\documentclass[twocolumn]{jpsj3}

\usepackage{color}
\usepackage{bm}

\title{%
Effective Model for Massless Dirac Electrons on a Surface of
Weak Topological Insulators
}

\author{%
Takashi Arita and Yositake
Takane\thanks{takane@hiroshima-u.ac.jp}
}

\inst{%
Department of Quantum Matter, Graduate School of Advanced Sciences of Matter,\\
Hiroshima University, Higashihiroshima, Hiroshima 739-8530, Japan
}

\recdate{ \hspace{50mm} }

\abst{%
In a typical situation, gapless surface states of a three-dimensional (3D)
weak topological insulator (WTI) appear only on the sides,
leaving the top and bottom surfaces gapped.
To describe massless Dirac electrons emergent on such side surfaces of a WTI,
a two-dimensional (2D) model consisting of
a series of one-dimensional helical channels is usually employed.
However, an explicit derivation of such a model
from a 3D bulk Hamiltonian has been lacking.
Here, we explicitly derive an effective 2D model for the WTI surface states
starting from the Wilson-Dirac Hamiltonian for the bulk WTI
and establish a firm basis for the hitherto hypothesized 2D model.
We show that the resulting 2D model accurately reproduces
the excitation spectrum of surface Dirac electrons determined by the 3D model.
We also show that the 2D model is applicable to a side surface
with atomic steps.
}

\begin{document}
\sloppy
\maketitle

\section{Introduction}

Three-dimensional (3D) weak topological insulators (WTIs) are known to be
equivalent to stacked layers of two-dimensional (2D) quantum spin-Hall (QSH)
insulators.~\cite{fu,moore,roy}
The stacking direction is specified by the weak vector
$\mib{\nu} \equiv (\nu_1,\nu_2,\nu_3)$, where $\nu_1$, $\nu_2$, and $\nu_3$
are called weak indices.
Reflecting the feature of a 2D QSH insulator that it is gapped in its bulk
but possesses a gapless one-dimensional (1D) helical channel
at its edge,\cite{kane,bernevig}
a WTI accommodates low-energy electron states arising from
helical edge channels only on its side surface.
This should be contrasted to the case of strong topological insulators,
in which low-energy electron states appear on every surface.
We refer to low-energy surface electrons as Dirac electrons
since they obey the massless Dirac equation.
A characteristic feature of WTIs is that their low-energy surface states
typically consist of two Dirac cones in the reciprocal space,
in contrast to the case of strong topological insulators,
where typically only one Dirac cone is present.
Owing to this, the surface state of a WTI was considered to be weak
against disorder, becoming gapped by scattering between two Dirac cones.
However, it has been shown that a WTI is not necessarily weak.~\cite{ran,
imura1,ringel,mong,liu1,imura2,yoshimura,kobayashi,morimoto,obuse,takane}
As low-energy electron states on the side surface of a WTI
are formed by a series of helical edge channels,
they are significantly affected by whether the number of QSH layers
stacked along $\mib{\nu}$ is even or odd.~\cite{ringel,imura2,yoshimura}
If it is even, the helical edge channels acquire a finite-size gap
owing to their mutual coupling.
Contrastingly, if it is odd, one helical channel survives
and the system has a gapless excitation spectrum.
This parity dependence is another characteristic feature
of surface Dirac electrons.
Several materials have been proposed
as possible WTIs.~\cite{yan,rasche,tang,yang}

To theoretically describe Dirac electrons on a side surface of WTIs,
an effective 2D model consisting of coupled 1D helical channels
has been proposed in Refs.~\citen{morimoto} and \citen{obuse}.
This model has two Dirac cones in the reciprocal space and is capable of
describing the even-odd parity dependence of an excitation spectrum
with respect to the number of QSH layers constituting a sample.
However, the connection between such an effective 2D model for surface states
and a 3D model for bulk WTIs has not been established concretely.

In this paper, we derive an effective 2D Hamiltonian for Dirac electrons
on a side surface of WTIs starting from the 3D Wilson-Dirac Hamiltonian
for bulk topological insulators.
We show that the resulting 2D model is indeed equivalent to
coupled 1D helical channels and
that all parameters in it are determined by those of the original 3D model.
By comparing the excitation spectrum of surface Dirac electrons obtained from
the 2D model with that obtained from the 3D model,
we confirm the validity of the effective model.
We also show that the effective model is applicable to
a side surface of WTIs with atomic steps.
We set $\hbar = 1$ throughout this paper.

\section{Derivation of the 2D Model}

We start from the following Wilson-Dirac Hamiltonian
for 3D topological insulators in the continuum limit:~\cite{liu2}
\begin{align}
   H
   = \epsilon_{\mib{k}}\mib{1} +
     \left[ 
       \begin{array}{cccc}
         M_{\mib{k}} & Bk_{z}a & 0 & Ak_{-}a \\
         Bk_{z}a & -M_{\mib{k}} & Ak_{-}a & 0 \\
         0 & Ak_{+}a & M_{\mib{k}} & -Bk_{z}a \\
         Ak_{+}a & 0 & -Bk_{z}a & -M_{\mib{k}}
       \end{array}
     \right] ,
\end{align}
where $\mib{1}$ is the $4 \times 4$ unit matrix, $a$ is the lattice constant,
$k_{\pm} = k_{x}\pm i k_{y}$, and
\begin{align}
  \epsilon_{\mib{k}}
  & = c_0 + c_{2\parallel}(k_{x}^{2}+k_{y}^{2})a^{2}
          + c_{2\perp}k_{z}^{2}a^{2} ,
        \\
  M_{\mib{k}}
  & = m_0 + m_{2\parallel}(k_{x}^{2}+k_{y}^{2})a^{2}
          + m_{2\perp}k_{z}^{2}a^{2} .
\end{align}
The basis set $\left\{|\mib{k}\rangle_{1\uparrow},|\mib{k}\rangle_{2\uparrow},
|\mib{k}\rangle_{1\downarrow},|\mib{k}\rangle_{2\downarrow} \right\}$
is adopted in expressing $H$ in the matrix form,
where $\uparrow, \downarrow$ and $1, 2$ respectively
represent the spin and orbital degrees of freedom.

\begin{figure}[btp]
\begin{center}
\includegraphics[height=3.0cm]{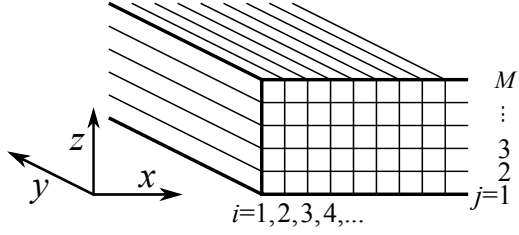}
\end{center}
\caption{Structure of a WTI sample considered in the text;
it consists of $M$ layers in the $z$-direction with $M \ge j \ge 1$
and is semi-infinite in the $x$-direction with $i \ge 1$,
while it is infinitely long in the $y$-direction.
}
\end{figure}
Among several topological phases described by this model
in a discretized version,~\cite{imura2} we choose the weak topological phase
with $\mib{\nu} \equiv (0,0,1)$ as a typical example.
In this phase, the system is equivalent to 2D QSH insulators
stacked in the $z$-direction.
We discretize the $z$- and $x$-coordinates and implement $H$
on the square lattice on the $xz$-plane leaving the $y$-coordinate unchanged.
The discretization of the $z$-coordinate is indispensable
to take account of the parity effect
that depends on whether the number of QSH layers is even or odd.
Let us use the indices $i$ and $j$ respectively to specify lattice sites
in the $x$- and $z$-directions.
We assume that the system consists of $M$ layers in the $z$-direction
with $M \ge j \ge 1$
and is semi-infinite in the $x$-direction with $i \ge 1$ (see Fig.~1).
We also assume that the system is infinitely long in the $y$-direction,
so $k_y$ remains a good quantum number.
Our attention is focused on Dirac electrons emerging on the side surface
of height $M$ in the $yz$-plane.
Let us introduce the four-component state vector for the $(i,j)$th site,
\begin{align}
  |i,j \rangle
  =  \left[ |i,j \rangle_{1\uparrow},
             |i,j \rangle_{2\uparrow},
             |i,j \rangle_{1\downarrow},
             |i,j \rangle_{2\downarrow}
     \right] .
\end{align}
In terms of this vector,
the discretized version of the Wilson-Dirac Hamiltonian is expressed as 
\begin{align}
   H_{\rm 3D} = H_{x}+H_{y}+H_{z}
\end{align}
with
\begin{align}
   H_{x}
 & = \sum_{i=1}^{\infty}\sum_{j=1}^{M}
     \Bigl[ |i,j \rangle h_{0} \langle i,j|
            + \bigl\{ |i+1,j \rangle h_{x}^{+} \langle i,j|
                      + {\rm h.c.}
              \bigr\}
     \Bigr] ,
         \\
   H_{y}
 & = \sum_{i=1}^{\infty}\sum_{j=1}^{M}
     |i,j \rangle h_{y} \langle i,j| ,
         \\
   H_{z}
 & = \sum_{i=1}^{\infty}\sum_{j=1}^{M-1}
     \left\{ |i,j+1 \rangle h_{z}^{+} \langle i,j|
             + {\rm h.c.} \right\} .
\end{align}
Here, the $4 \times 4$ matrices are given by
\begin{align}
   h_{0}
 & = \tilde{c}_{0}\mib{1} +
     \left[ 
       \begin{array}{cccc}
         \tilde{m}_{0} & 0 & 0 & 0 \\
         0 & -\tilde{m}_{0} & 0 & 0 \\
         0 & 0 & \tilde{m}_{0} & 0 \\
         0 & 0 & 0 & -\tilde{m}_{0}
       \end{array}
     \right] ,
               \\
   h_{x}^{+}
 & = -c_{2\parallel}\mib{1} +
     \left[ 
       \begin{array}{cccc}
         -m_{2\parallel} & 0 & 0 & \frac{i}{2}A \\
         0 & m_{2\parallel} & \frac{i}{2}A & 0 \\
         0 & \frac{i}{2}A & -m_{2\parallel} & 0 \\
         \frac{i}{2}A & 0 & 0 & m_{2\parallel}
       \end{array}
     \right] ,
               \\
   h_{y}
 & = \left[ 
       \begin{array}{cccc}
         \xi_{+}(k_y) & 0 & 0 & -iAk_{y}a \\
         0 & \xi_{-}(k_y) & -iAk_{y}a & 0 \\
         0 & iAk_{y}a & \xi_{+}(k_y) & 0 \\
         iAk_{y}a & 0 & 0 & \xi_{-}(k_y)
       \end{array}
     \right] ,
               \\
   h_{z}^{+}
 & = -c_{2\perp}\mib{1} +
     \left[ 
       \begin{array}{cccc}
         -m_{2\perp} & \frac{i}{2}B & 0 & 0 \\
         \frac{i}{2}B & m_{2\perp} & 0 & 0 \\
         0 & 0 & -m_{2\perp} & -\frac{i}{2}B \\
         0 & 0 & -\frac{i}{2}B & m_{2\perp}
       \end{array}
     \right] ,
\end{align}
where
\begin{align}
    \tilde{c}_{0}
 & = c_{0}+2c_{2\parallel}+2c_{2\perp} ,
               \\
    \tilde{m}_{0}
 & = m_{0}+2m_{2\parallel}+2m_{2\perp} ,
               \\
   \xi_{\pm}(k_y) 
 & = \left(c_{2\parallel}\pm m_{2\parallel}\right) k_{y}^{2}a^{2} .
\end{align}
Note that the Wilson mass term $M_{\mib{k}}$ in the original continuum model
is now modified to
\begin{align}
  M_{\mib{k}}^{\rm dis}
  & = m_0 + m_{2\parallel}\left\{2[1-\cos(k_{x}a)]+(k_{y}a)^{2}\right\}
      \nonumber \\
  & \hspace{5mm}
    + m_{2\perp}2[1-\cos(k_{z}a)] .
\end{align}
We focus on the weak topological phase with $\mib{\nu} \equiv (0,0,1)$
stabilized when the parameters satisfy~\cite{imura2}
\begin{align}
    \label{eq:condition-WTI}
  m_{2\parallel} > \frac{1}{4}|m_0| > m_{2\perp}
                > \frac{1}{4}|m_0| - m_{2\parallel} ,
\end{align}
where $m_{0} < 0$ and $m_{2\parallel} > |c_{2\parallel}| \ge 0$ are assumed.
This condition fixes the sign of the mass term at four symmetric points
on the $k_{x}k_{z}$-plane with $k_{y}=0$ as follows:
\begin{align}
  M_{\mib{k}}^{\rm dis}
  = \left\{ \begin{array}{ll}
               m_{0}<0,
                & \mib{k}' = (0,0) \\
               m_{0}+4m_{2\parallel}>0,
                & \mib{k}' = (\pi/a,0) \\
               m_{0}+4m_{2\perp}<0,
                & \mib{k}' = (0,\pi/a) \\
               m_{0}+4m_{2\parallel}+4m_{2\perp}>0,
                & \mib{k}' = (\pi/a,\pi/a)
            \end{array}
    \right. , 
\end{align}
where $\mib{k}' = (k_{x},k_{z})$.
This indicates that, on the side surface in the $yz$-plane,
the Dirac point appears at $(k_{y},k_{z}) = (0,0)$ and $(0,\pi/a)$.

Let us find two basis functions for low-energy states localized
near the surface by solving the eigenvalue equation
for the $x$-direction with $j$ fixed.
We show below that the resulting basis functions
describe the 1D helical channel arising from the $j$th QSH layer.
The procedure is similar to that of Ref.~\citen{okamoto}
developed on the basis of earlier works.~\cite{liu2,konig,shan}
The eigenvalue equation now of concern is written as
\begin{align}
      \label{eq:EE-x}
   H_{x}|\psi(j)\rangle = E_{\perp}|\psi(j)\rangle
\end{align}
for a given $j$, where
\begin{align}
      \label{eq:def-psi_ket}
    |\psi(j)\rangle = \sum_{i=1}^{\infty}|i,j\rangle\mib{\psi}(i) .
\end{align}
Since its solutions localized near the surface are necessary for our argument,
the appropriate boundary condition for $\mib{\psi}(i)$ is
$\mib{\psi}(0) = \mib{\psi}(\infty) = {}^{t}(0,0,0,0)$.
Solving Eq.~(\ref{eq:EE-x}) under the required boundary condition,
we obtain two degenerate solutions
$|\psi_{+}(j)\rangle$ and $|\psi_{-}(j)\rangle$ with
\begin{align}
     \label{eq:res-E_perp}
  E_{\perp}
  = \tilde{c}_{0}-\frac{c_{2\parallel}}{m_{2\parallel}}\tilde{m}_{0} ,
\end{align}
where the detailed derivation is given in the Appendix.
The resulting expression of $|\psi_{\pm}(j)\rangle$ is given as
\begin{align}
     \label{eq:psi-pm_ket}
  |\psi_{\pm}(j)\rangle = |\psi_{0}(j)\rangle \mib{v}_{\pm}
\end{align}
with
\begin{align}
  \mib{v}_{+}
  & = \frac{1}{\sqrt{2}}
      \left[ \begin{array}{c}
               0 \\
               -i\sqrt{1+\frac{c_{2\parallel}}{m_{2\parallel}}} \\
               \sqrt{1-\frac{c_{2\parallel}}{m_{2\parallel}}} \\
               0
             \end{array}
      \right] ,
         \\
  \mib{v}_{-}
  & = \frac{1}{\sqrt{2}}
      \left[ \begin{array}{c}
               \sqrt{1-\frac{c_{2\parallel}}{m_{2\parallel}}} \\
               0 \\
               0 \\
               -i\sqrt{1+\frac{c_{2\parallel}}{m_{2\parallel}}}
             \end{array}
      \right] ,
\end{align}
and
\begin{align}
     \label{eq:psi_0-j}
   |\psi_{0}(j)\rangle
   = \mathcal{C} \sum_{i=1}^{\infty}
     \left(\rho_{+}^{i}-\rho_{-}^{i}\right)|i,j \rangle ,
\end{align}
where $\mathcal{C}$ is a normalization constant, and
$\rho_{+}$ and $\rho_{-}$ are constants given by Eq.~(\ref{eq:rho-pm})
satisfying $|\rho_{\pm}| < 1$.
Note that $|\psi_{+}(j)\rangle$ and $|\psi_{-}(j)\rangle$ play
the role of the basis functions for low-energy surface states.
Clearly, $|\psi_{0}(j)\rangle$ represents the penetration of
surface states into the bulk.

In terms of the basis functions presented above,
we can express a low-energy surface state as
\begin{align}
  |\Psi\rangle
  = \sum_{j=1}^{M} \bigl( \alpha_{j}|\psi_{+}(j)\rangle
                          + \beta_{j}|\psi_{-}(j)\rangle \bigr) .
\end{align}
We derive an effective Hamiltonian for $\alpha_{j}$ and $\beta_{j}$
in the following.~\cite{liu2}
For this state, the eigenvalue equation is written as
\begin{align}
     \label{eq:eigen-eq_red}
  \left(H_{y}+H_{z}\right)|\Psi\rangle
  = \left(E-E_{\perp}\right)|\Psi\rangle .
\end{align}
Taking the inner product of both sides of Eq.~(\ref{eq:eigen-eq_red})
with $\mib{v}_{+}^{\dagger}\langle\psi_{0}(j)|$
and $\mib{v}_{-}^{\dagger}\langle\psi_{0}(j)|$,
we obtain a set of equations for $\alpha_{j}$ and $\beta_{j}$:
\begin{align}
      \label{eq:red-1}
  & \gamma A(k_{y}a)\alpha_{j} + t\alpha_{j-1} + t\alpha_{j+1}
        \nonumber \\
  & \hspace{7mm}
    - \frac{\gamma}{2}B\beta_{j-1} + \frac{\gamma}{2}B\beta_{j+1}
    = \left(E-E_{\perp}\right)\alpha_{j} ,
      \\
      \label{eq:red-2}
  & -\gamma A(k_{y}a)\beta_{j} + t\beta_{j-1} + t\beta_{j+1}
        \nonumber \\
  & \hspace{7mm}
    + \frac{\gamma}{2}B\alpha_{j-1} - \frac{\gamma}{2}B\alpha_{j+1}
    = \left(E-E_{\perp}\right)\beta_{j} ,
\end{align}
where
\begin{align}
      \label{eq:res-gamma}
   \gamma = \sqrt{1-\left(\frac{c_{2\parallel}}{m_{2\parallel}}\right)^{2}} ,
        \\
      \label{eq:res-tz}
   t 
   = -c_{2\perp} + \frac{c_{2\parallel}}{m_{2\parallel}}m_{2\perp} .
\end{align}
Let us rewrite the basis functions as
$|\psi_{+}(j)\rangle \to |j\rangle_{\uparrow}$ and
$|\psi_{-}(j)\rangle \to |j\rangle_{\downarrow}$,
erasing the degree of freedom with respect to the $x$-direction,
and express an arbitrary wave function as
\begin{align}
  |\Psi\rangle
  = \sum_{j=1}^{M} |j\rangle
                   \left[ \begin{array}{c}
                            \alpha_j \\ \beta_j
                          \end{array}
                   \right]
\end{align}
with
$|j\rangle \equiv \left\{|j\rangle_{\uparrow},|j\rangle_{\downarrow}\right\}$.
Then, the effective 2D Hamiltonian is given by
\begin{align}
       \label{H_2D}
   H_{\rm 2D}
 & = \sum_{j=1}^{M}
     |j\rangle \left[ \begin{array}{cc}
                          E_{\perp}+\gamma Ak_{y}a & 0 \\
                          0 & E_{\perp}-\gamma Ak_{y}a
                      \end{array} \right]
     \langle j|
   \nonumber \\
 &  \hspace{0mm}
   + \sum_{j=1}^{M-1}
     \left\{
     |j+1\rangle \left[ \begin{array}{cc}
                          t & -\frac{\gamma}{2}B \\
                          \frac{\gamma}{2}B & t
                          \end{array} \right]
     \langle j|
   + {\rm h.c.}
     \right\} .
\end{align}
We can show that $H_{\rm 2D}|\Psi\rangle=E|\Psi\rangle$ is
equivalent to Eqs.~(\ref{eq:red-1}) and (\ref{eq:red-2}).
Clearly, $|j\rangle_{\uparrow}$ and $|j\rangle_{\downarrow}$
respectively represent the right-going and left-going branches of
the edge helical channel arising from the $j$th QSH layer.
This indicates that the derived model is equivalent to a series of
1D helical channels, each of which is coupled with its nearest neighbors.
The expression of $H_{\rm 2D}$ with Eqs.~(\ref{eq:res-E_perp}),
(\ref{eq:res-gamma}), and (\ref{eq:res-tz})
is the central result of this paper.

\section{Analytical Treatment of the 2D Model}

Let us briefly consider the effective 2D Hamiltonian
given in Eq.~(\ref{H_2D}) in an analytical manner.
If the periodic boundary condition is artificially imposed
in the $z$-direction and the limit of $M \to \infty$ is taken,
the dispersion relation is given by
\begin{align}
  E = E_{\perp} + 2t\cos k_{z}a
      \pm \gamma\sqrt{(Ak_{y}a)^{2}+(B\sin k_{z}a)^{2}} .
\end{align}
This indicates that two Dirac cones centered at $(k_{y},k_{z})=(0,0)$
and $(0,\pi/a)$ appear in the reciprocal space.~\cite{imura2}
This is a characteristic feature of WTIs.
The Dirac point energies $E_{0}$ at $(0,0)$ 
and $E_{\pi}$ at $(0,\pi/a)$ are respectively expressed as
\begin{align}
  E_{0}
 & = E_{\perp} + 2t
       \nonumber \\
 & =  c_{0}+2c_{2\parallel}
     -\frac{c_{2\parallel}}{m_{2\parallel}}\left(m_{0}+2m_{2\parallel}\right) ,
                 \\
  E_{\pi}
 & = E_{\perp} - 2t
       \nonumber \\
 & = c_{0}+2c_{2\parallel}+4c_{2\perp}
     -\frac{c_{2\parallel}}{m_{2\parallel}}
      \left(m_{0}+2m_{2\parallel}+4m_{2\perp}\right) .
\end{align}

Turning to the realistic case in which Dirac electrons are confined
in the finite region of $M \ge j \ge 1$,
we construct eigenstates at an energy $\epsilon$.
In this case, a subband structure should appear
reflecting the confinement of Dirac electrons.
It is worth mentioning that Dirac electrons cannot be confined
if only one Dirac cone exists in the reciprocal space.
The presence of two Dirac cones enables the confinement as we see below.
For simplicity, we restrict our consideration to the case of $t = 0$.
Let us assume that eigenfunctions are expressed in the form of
\begin{align}
     \label{eq:def-two-vect}
   \left[ \begin{array}{c}
             \alpha_j \\ \beta_j
          \end{array}
   \right]
   = \chi(j)
     \left[ \begin{array}{c}
              a \\
              b
            \end{array}
     \right] ,
\end{align}
where the transverse function $\chi(j)$ must satisfy
the boundary condition of $\chi(0)=\chi(M+1)=0$.
We can construct $\chi(j)$ that satisfies this condition
by superposing two wave functions of different Dirac cones~\cite{imura2}
sharing an identical eigenvector.
As a result, we find that
\begin{align}
    \label{eq:transv-f}
  \chi_{m}(j)
  & \propto \left(e^{iq_{m}z_j}
                  -e^{i\left(\frac{\pi}{a}-q_{m}\right)z_j} \right)
         \nonumber \\
  & \propto \left(e^{iq_{m}z_j}-(-1)^{j}e^{-iq_{m}z_j} \right) ,
\end{align}
where $z_{j} = ja$ and $q_{m} = m\pi/[(M+1)a]$ with
\begin{align}
     \label{eq:m-even}
  m = \pm \frac{1}{2}, \pm \frac{3}{2}, \dots, \pm \frac{M-1}{2}
\end{align}
for an even $M$, and
\begin{align}
  m = 0, \pm 1, \pm 2, \dots, \pm \frac{M-1}{2}
\end{align}
for an odd $M$.
The dispersion relation for the $m$th subband is given by
\begin{align}
     \label{eq:dis-analytic}
  E_{m}(k_{y})
  = E_{\perp} \pm \sqrt{(\gamma Ak_{y}a)^{2}+\Delta_{m}^2}
\end{align}
with
\begin{align}
  \Delta_{m} = \gamma B \sin q_{m}a .
\end{align}
The corresponding eigenvector is expressed as
\begin{align}
   \left[ \begin{array}{c}
            a_{m} \\
            b_{m}
          \end{array} \right]
   \propto
   \left[ \begin{array}{c}
            -i\Delta_{m} \\
            E_{\perp}+\gamma Ak_{y}a-E_{m}(k_{y})
          \end{array} \right] .
\end{align}
Each subband with $m \neq 0$ is doubly degenerate
since $|\Delta_{m}| = |\Delta_{-m}|$.
We see in the next section that this degeneracy is lifted when $t \neq 0$.

The parity effect can be observed in Eq.~(\ref{eq:dis-analytic}).
For an odd $M$, we see that $\Delta_{m}$ vanishes for $m = 0$,
indicating that the system has a gapless excitation spectrum.
Contrastingly, $m = 0$ is not allowed for an even $M$
as indicated in Eq.~(\ref{eq:m-even}),
so the finite-size gap $2\Delta_{\frac{1}{2}}$ opens across the Dirac point
at which $E = E_{\perp}$.
Here, it is meaningful to point out a peculiar property of
the zero-energy mode with $m = 0$.
Note that $\chi_{0}(j)$ given above has a finite amplitude
only at sites with an odd $j$ and vanishes otherwise.
Thus, even though disorder is introduced in $B$, this mode remains
an eigenfunction of $H_{\rm 2D}$ with $\Delta_{0} = 0$
although the corresponding vector $^{t}[a,b]$ in Eq.~(\ref{eq:def-two-vect})
is no longer $j$-independent.
In this sense, the zero-energy mode is robust against disorder.

\section{Comparison between the 2D and 3D Models}

\begin{figure}[btp]
\begin{center}
\includegraphics[height=2.5cm]{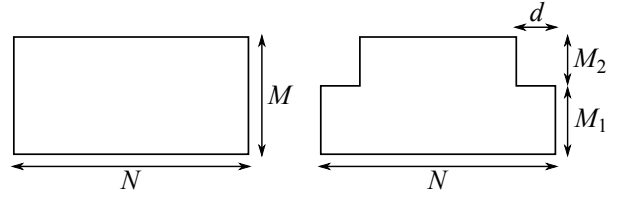}
\end{center}
\caption{Cross sections of prism-shaped systems considered in the text:
rectangular cross section (left)
and cross section with an atomic step of depth $d$
on both side surfaces (right).
}
\end{figure}
In this section, we numerically obtain an excitation spectrum
(i.e., subband structure) of Dirac electrons on a side surface of height $M$
on the basis of the effective 2D Hamiltonian $H_{\rm 2D}$.
By comparing the resulting subband structure with that obtained from
the 3D bulk Hamiltonian $H_{\rm 3D}$,
we examine the validity of our effective model.
In determining the subband structure on the basis of $H_{\rm 3D}$,
we consider an infinitely long rectangular prism-shaped system of height $M$
and width $N$ (i.e., $N \ge i \ge 1$).
Its cross section is shown in the left panel of Fig.~2.
This system has the two side surfaces, on which low-energy states appear.
We expect that, if $N$ is chosen to be sufficiently large,
the low-energy states on one surface and those on the other surface are
exactly degenerate without mutual coupling,
and that both of them are comparable to those described by $H_{\rm 2D}$.
Note that $H_{\rm 3D}$ provides us with not only the subband structure
of surface states but also the band structure of bulk states.
Setting $N = 20$, we perform numerical calculations for the two cases of
$t/A = 0$ and $t/A = 0.02$ to observe the effect of $t$.
The parameters are fixed except for $c_{2\perp}$ as follows:
$B/A = 0.4$, $m_0/A = -0.5$, $m_{2\parallel}/A = 0.5$,
$m_{2\perp}/A = -0.1$, $c_0/A =-1.0$, and $c_{2\parallel}/A = 0.02$.
The value of $c_{2\perp}$ is chosen as
$c_{2\perp}/A = -0.004$ in the case of $t/A = 0$
and $c_{2\perp}/A = -0.024$ in the case of $t/A = 0.02$.
We find that $E_{\perp}/A = -0.98$ in the case of $t/A = 0$
and $E_{\perp}/A = -1.02$ in the case of $t/A = 0.02$.

\begin{figure}[tbp]
\begin{center}
\includegraphics[height=5.5cm]{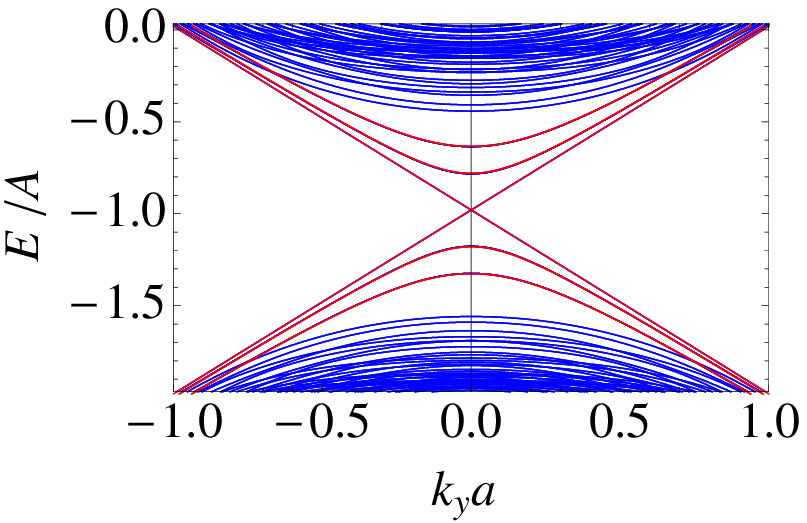}
\includegraphics[height=5.5cm]{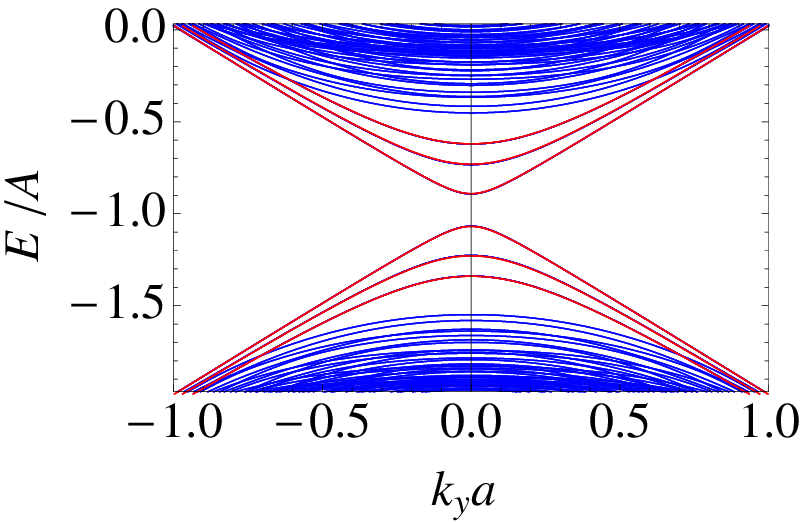}
\end{center}
\caption{(Color online)
Band structures in the case of $t/A = 0$ for $M = 5$ (upper panel)
and $M = 6$ (lower panel), where dashed (red) lines and solid (blue) lines
respectively represent the results
obtained from the effective 2D model and 3D model.
}
\end{figure}
\begin{figure}[tbp]
\begin{center}
\includegraphics[height=5.5cm]{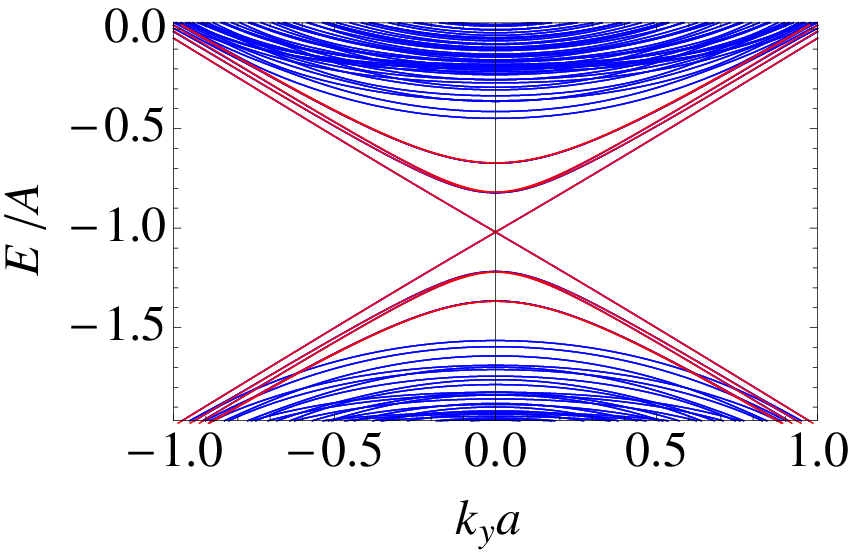}
\includegraphics[height=5.5cm]{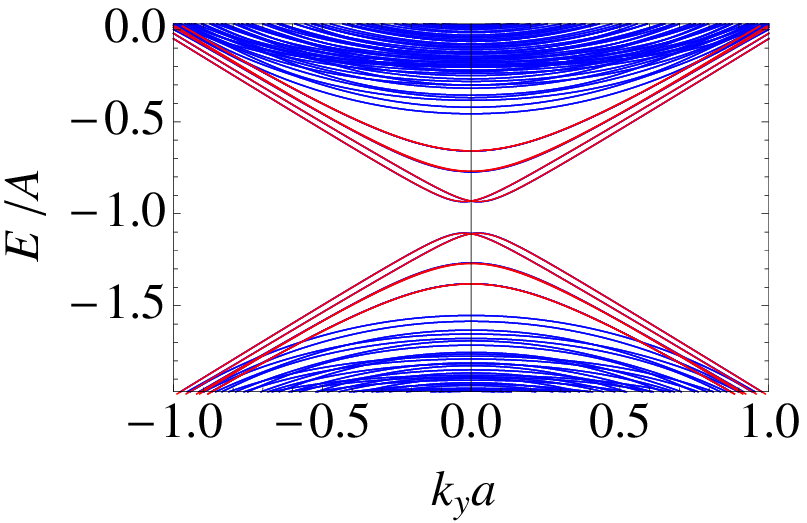}
\end{center}
\caption{(Color online)
Band structures in the case of $t/A = 0.02$ for $M = 5$ (upper panel)
and $M = 6$ (lower panel), where dashed (red) lines and solid (blue) lines
respectively represent the results
obtained from the effective 2D model and 3D model.
}
\end{figure}
Figure~3 shows the band structure in the case of $t/A = 0$
for $M = 5$ and $6$,
where dashed (red) lines and solid (blue) lines respectively represent
the results obtained from $H_{\rm 2D}$ and $H_{\rm 3D}$.
In the subband structure of surface states inside the bulk gap,
the solid (blue) lines completely overlap the dashed (red) ones.
This means that the result of the effective 2D model
is identical to that of the 3D model concerning the surface states.
As noted in the previous section, each subband of surface states
is doubly degenerate except for the one with a linear dispersion.
The band structure of bulk states consists of only
solid (blue) lines obtained from $H_{\rm 3D}$
since $H_{\rm 2D}$ can describe only surface states.
Figure~4 shows the band structure in the case of $t/A = 0.02$
for $M = 5$ and $6$.
Again, the solid (blue) lines completely overlap
the dashed (red) ones for surface states,
indicating that the result of the effective 2D model is
identical to that of the 3D model.
In this case, the degeneracy of each subband of surface states
is slightly lifted due to the nonzero $t$.

In the remainder of this section,
we examine the applicability of the effective 2D model
to the situation where a side surface contains an atomic step.
As an example, let us consider a prism-shaped system
with the cross section shown in the right panel of Fig.~2,
where both side surfaces of the $M$ layers ($M \equiv M_{1}+M_{2}$)
contain an atomic step of depth $d$.
We assume that the left and right atomic steps are located between
the $M_{1}$th layer and the $M_{1}+1$th layer, separating the system
into subsystems of $M_{1}$ and $M_{2}$ layers.
Both the side surfaces have an identical subband structure,
which is determined by $H_{\rm 3D}$.
Our purpose is to answer the question of whether the subband structure
can be reproduced on the basis of $H_{\rm 2D}$.
The most important effect of an atomic step upon surface electrons is that
it reduces electron hopping between neighboring helical channels across it.
Thus, we expect to be able to describe the subband structure
in terms of $H_{\rm 2D}$ by reducing the corresponding hopping terms.
According to this observation, we propose the use of the following Hamiltonian:
\begin{align}
   \tilde{H}_{\rm 2D}
 & = \sum_{j=1}^{M}
     |j\rangle \left[ \begin{array}{cc}
                          E_{\perp}+\gamma Ak_{y}a & 0 \\
                          0 & E_{\perp}-\gamma Ak_{y}a
                      \end{array} \right]
     \langle j|
   \nonumber \\
 &  \hspace{-4mm}
   + \sum_{j=1}^{M-1} \eta_{j}
     \left\{
     |j+1\rangle
                 \left[ \begin{array}{cc}
                          t & -\frac{\gamma}{2}B \\
                          \frac{\gamma}{2}B & t
                          \end{array} \right]
     \langle j|
   + {\rm h.c.}
     \right\} ,
\end{align}
where
\begin{align}
   \eta_{j}
   = \left\{ \begin{array}{cc}
                \delta & (j = M_{1}) \\
                1 & (j \neq M_{1})
             \end{array}
     \right.
\end{align}
with $\delta < 1$.
As $\delta$ represents the overlap between the basis functions
for the $M_{1}$th and $M_{1}+1$th helical channels,
we expect from Eq.~(\ref{eq:psi_0-j}) that
it decreases roughly exponentially with increasing $d$.
Treating $\delta$ as a fitting parameter, we try to reproduce
the subband structure using $\tilde{H}_{\rm 2D}$.

\begin{figure}[tbp]
\begin{center}
\includegraphics[height=5.5cm]{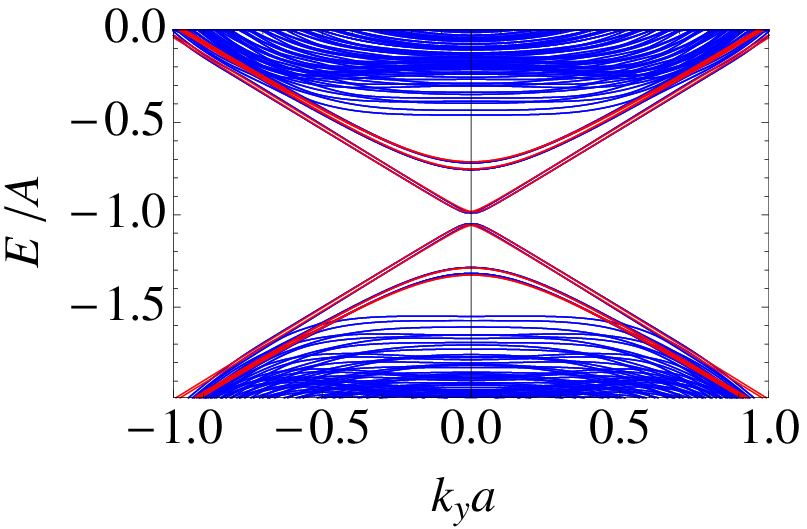}
\includegraphics[height=5.5cm]{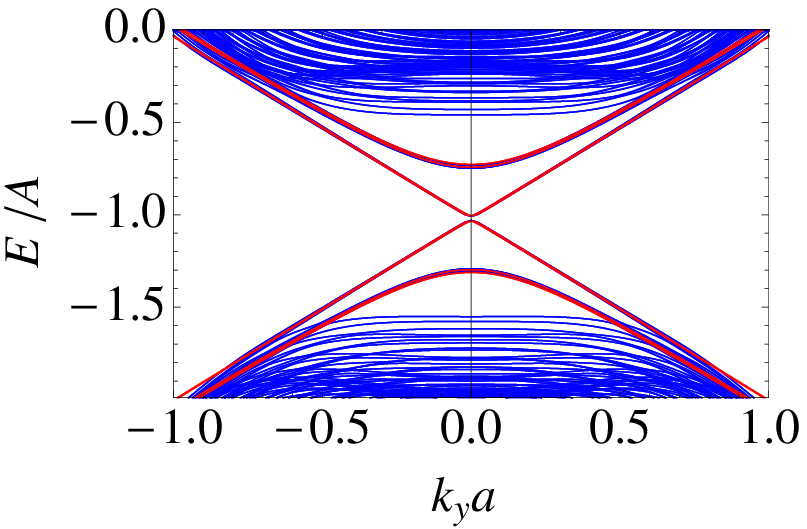}
\includegraphics[height=5.5cm]{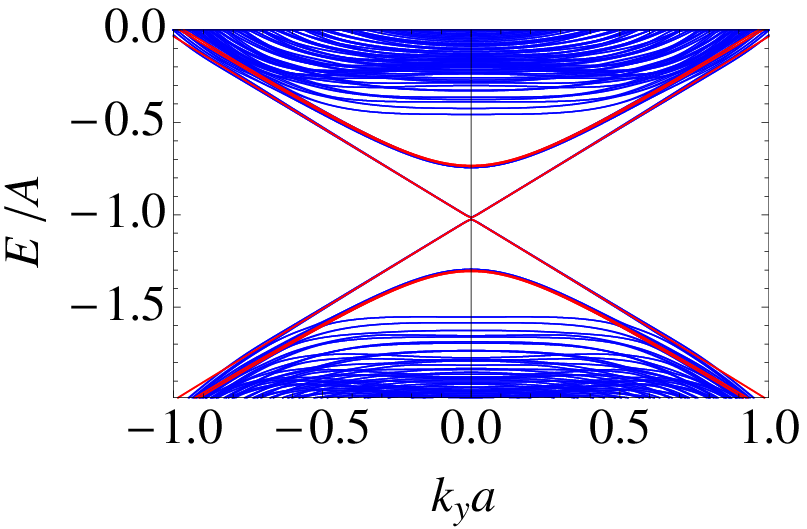}
\end{center}
\caption{(Color online)
Band structures in the case of $M_1 = M_2 = 3$ for $d = 1$ (upper panel),
$d = 2$ (middle panel),  and $d = 3$ (lower panel), where dashed (red) lines
and solid (blue) lines respectively represent the results obtained from
the effective 2D model and 3D model.
}
\end{figure}
Figure~5 shows the band structure in the case of $M_{1} = M_{2} = 3$
with step depths $d = 1$, $2$, and $3$ from top to bottom,
where dashed (red) lines and solid (blue) lines respectively represent
the results obtained from $\tilde{H}_{\rm 2D}$ and $H_{\rm 3D}$.
The set of parameters in the case of $t/A = 0.02$ is employed,
and the best fitting is found at
$\delta = 0.375$, $0.15$, and $0.05$, respectively, for $d = 1$, $2$, and $3$.
In Fig.~5, the solid (blue) lines again completely overlap
the dashed (red) ones in the subband structure of surface states
inside the bulk gap.
This means that $\tilde{H}_{\rm 2D}$ accurately reproduces
the subband structure even in the presence of an atomic step.

As long as we focus on surface states, the system under consideration can be
regarded as two coupled side surfaces consisting of three layers.
It is clearly equivalent to a six-layer side surface at $d = 0$,
and it approaches two decoupled side surfaces with increasing $d$.
We find from Fig.~5 that the lowest subband shows a nearly linear
dispersion, which is a characteristic feature of odd-layer systems.
We also find that the lowest subband reveals a finite-size gap
that very rapidly decreases with increasing $d$.
Obviously, the presence of a finite-size gap is
a characteristic feature of even-layer systems.
This implies that the subband structure of surface states in the presence of
an atomic step reflects not only the parity of $M_1$ and $M_2$
but also that of $M_1+M_2$ when $d$ is sufficiently small.

\section{Summary}

In this paper, we have derived an effective 2D model for Dirac electrons
on a side surface of weak topological insulators
starting from the 3D Wilson-Dirac Hamiltonian for bulk topological insulators.
Although the resulting 2D model itself is similar to those proposed
in Refs.~\citen{morimoto} and \citen{obuse},
our approach has an advantage that all the parameters in the 2D model are
directly connected with those in the original 3D model.
It is shown that the 2D model accurately reproduces the spectrum of
surface Dirac electrons determined by the 3D model, indicating its validity.
It is also shown that the model is applicable to a side surface
with an atomic step.
Although only the case with a single step is treated there,
the extension to a case with multiple steps is straightforward.

\section*{Acknowledgment}

The authors thank K.-I. Imura and Y. Yoshimura for valuable discussions.
This work was supported by a Grant-in-Aid for Scientific Research (C)
(No. 24540375).

\appendix

\section{}

Let us find two wave functions $\mib{\psi}(j)$ with which
$|\psi(j)\rangle$ satisfies the eigenvalue equation~(\ref{eq:EE-x}).
Considering the matrix form of $h_{x}^{+}$, we find that
one of them is in the form of $\mib{\psi}(j) = {}^{t}(v_1,0,0,v_2)$
and the other is $\mib{\psi}(j) = {}^{t}(0,v_2,v_1,0)$.
Below, we mainly treat the first type, $\mib{\psi}(j) = {}^{t}(v_1,0,0,v_2)$,
as the second type can be obtained
by rearranging the elements of the first type.

Let us find elementary solutions $\mib{\phi}(i)$ of
the eigenvalue equation~(\ref{eq:EE-x})
assuming $\mib{\phi}(i) = \rho^{i}\mib{v}$
with $\mib{v}={}^{t}(v_1,0,0,v_2)$.
Under this assumption, the eigenvalue equation is reduced to
\begin{align}
     \label{eq:red-eigen_V}
  \left[
    \begin{array}{cccc}
      \epsilon(\rho)+M(\rho) & A(\rho) \\
      A(\rho) & \epsilon(\rho)-M(\rho)
    \end{array}
  \right] \mib{v}^{\prime}
 & = E_{\perp} \mib{v}^{\prime} ,
\end{align}
where $\mib{v}^{\prime} = {}^{t}(v_1,v_2)$ and
\begin{align}
  \epsilon(\rho)
 & = \tilde{c}_{0} - c_{2\parallel}\left(\rho+\rho^{-1}\right) ,
        \\
  M(\rho)
 & = \tilde{m}_{0} - m_{2\parallel}\left(\rho+\rho^{-1}\right) ,
        \\
  A(\rho)
 & = -\frac{i}{2}A\left(\rho-\rho^{-1}\right) .
\end{align}
Equation~(\ref{eq:red-eigen_V}) holds only when
\begin{align}
       \label{eq:eigen-value}
  \left[\epsilon(\rho)-E_{\perp}\right]^{2}-M(\rho)^{2}-A(\rho)^{2} = 0 .
\end{align}
Let $\mib{\phi}_{\pm}(i) \equiv \rho_{\pm}^{i} \mib{v}_{\pm}$
be two different elementary solutions of Eq.~(\ref{eq:EE-x}),
in terms of which we can express a general solution as
\begin{align}
  \mib{\psi}(i)
  = d_{+} \rho_{+}^{i} \mib{v}_{+}
  + d_{-} \rho_{-}^{i} \mib{v}_{-} .
\end{align}
The boundary condition of $\mib{\psi}(\infty) = {}^{t}(0,0,0,0)$ requires
\begin{align}
  |\rho_{\pm}| < 1 .
\end{align}
The other boundary condition of $\mib{\psi}(0) = {}^{t}(0,0,0,0)$ requires
\begin{align}
  \mib{v}_{+} = \mib{v}_{-}
\end{align}
for $\rho_{+} \neq \rho_{-}$ with $d_{-} = - d_{+}$.

Now we consider the case when $\mib{v}_{+} = \mib{v}_{-}$
(or equivalently $\mib{v}_{+}^{\prime} = \mib{v}_{-}^{\prime}$) holds.
It is instructive to rewrite Eq.~(\ref{eq:red-eigen_V}) as
\begin{align}
  \left[
    \begin{array}{cccc}
      \frac{\epsilon(\rho_{\pm})+M(\rho_{\pm})-E_{\perp}}{A(\rho_{\pm})}
         & 1 \\
      1 & \frac{\epsilon(\rho_{\pm})-M(\rho_{\pm})-E_{\perp}}{A(\rho_{\pm})}
    \end{array}
  \right] \mib{v}^{\prime}_{\pm}
  =  \mib{0} .
\end{align}
This indicates that $\mib{v}_{+}^{\prime} = \mib{v}_{-}^{\prime}$
is realized for $\rho_{+} \neq \rho_{-}$ only when
\begin{align}
    \frac{\epsilon(\rho_{+})+M(\rho_{+})-E_{\perp}}{A(\rho_{+})}
    = \frac{\epsilon(\rho_{-})+M(\rho_{-})-E_{\perp}}{A(\rho_{-})}
\end{align}
and
\begin{align}
    \frac{\epsilon(\rho_{+})-M(\rho_{+})-E_{\perp}}{A(\rho_{+})}
    = \frac{\epsilon(\rho_{-})-M(\rho_{-})-E_{\perp}}{A(\rho_{-})}
\end{align}
simultaneously hold.~\cite{okamoto}
These equations require that
$M(\rho_{\pm}) \propto \epsilon(\rho_{\pm})-E_{\perp}$.
This relation determines $E_{\perp}$ as
\begin{align}
  E_{\perp}
  = \tilde{c}_{0}-\frac{c_{2\parallel}}{m_{2\parallel}}\tilde{m}_{0} ,
\end{align}
which can be rewritten as
\begin{align}
     \label{eq:ep-M}
  \epsilon(\rho_{\pm})-E_{\perp}
  = \frac{c_{2\parallel}}{m_{2\parallel}}M(\rho_{\pm}).
\end{align}
Combining Eqs.~(\ref{eq:eigen-value}) and (\ref{eq:ep-M}), we find that
\begin{align}
     \label{eq:M-A}
  \gamma M(\rho_{\pm}) = \pm i A(\rho_{\pm})
\end{align}
with
\begin{align}
   \gamma = \sqrt{1-\left(\frac{c_{2\parallel}}{m_{2\parallel}}\right)^{2}} .
\end{align}
As shown later, solutions with $|\rho_{\pm}| < 1$ are always obtained
in the case of $\gamma M(\rho_{\pm}) = i A(\rho_{\pm})$
under the condition of Eq.~(\ref{eq:condition-WTI}) with $A > 0$.
This immediately yields
\begin{align}
     \label{eq:rho-pm}
  \rho_{\pm}
  = \frac{\gamma\tilde{m}_{0}
          \pm\sqrt{(\gamma\tilde{m}_{0})^{2}-4(\gamma m_{2\parallel})^{2}
                   +A^{2}}}
         {2\left(\gamma m_{2\parallel}+\frac{A}{2}\right)}
\end{align}
and
\begin{align}
  \mib{v}^{\prime}
  = \frac{1}{\sqrt{2}}
    \left[ \begin{array}{c}
             \sqrt{1-\frac{c_{2\parallel}}{m_{2\parallel}}} \\
             -i\sqrt{1+\frac{c_{2\parallel}}{m_{2\parallel}}}
           \end{array}
    \right] ,
\end{align}
where $\mib{v}^{\prime} \equiv \mib{v}^{\prime}_{+}=\mib{v}^{\prime}_{-}$.
Now we can express the wave function $\mib{\psi}_{-}(i)$ of the first type as
\begin{align}
   \mib{\psi}_{-}(i)
   = \mathcal{C} \left(\rho_{+}^{i}-\rho_{-}^{i}\right)\mib{v}_{-}
\end{align}
with
\begin{align}
  \mib{v}_{-}
  = \frac{1}{\sqrt{2}}
    \left[ \begin{array}{c}
             \sqrt{1-\frac{c_{2\parallel}}{m_{2\parallel}}} \\
             0 \\
             0 \\
             -i\sqrt{1+\frac{c_{2\parallel}}{m_{2\parallel}}}
           \end{array}
    \right] ,
\end{align}
where $\mathcal{C}$ is a constant to be determined
by the normalization condition of
$\sum_{i=1}^{\infty}|\mathcal{C}\left(\rho_{+}^{i}-\rho_{-}^{i}\right)|^{2}=1$.
The wave function $\mib{\psi}_{+}(i)$ of the second type is expressed by
replacing $\mib{v}_{-}$ with $\mib{v}_{+}$ given by
\begin{align}
  \mib{v}_{+}
  = \frac{1}{\sqrt{2}}
    \left[ \begin{array}{c}
             0 \\
             -i\sqrt{1+\frac{c_{2\parallel}}{m_{2\parallel}}} \\
             \sqrt{1-\frac{c_{2\parallel}}{m_{2\parallel}}} \\
             0
           \end{array}
    \right] .
\end{align}
Substituting $\mib{\psi}_{\pm}(i)$ into Eq.~(\ref{eq:def-psi_ket}),
we obtain $|\psi_{\pm}(j)\rangle$ given in Eq.~(\ref{eq:psi-pm_ket})

Now we turn to Eq.~(\ref{eq:M-A})
and show that solutions with $|\rho_{\pm}| < 1$ are obtained only
in the case of $\gamma M(\rho_{\pm})=iA(\rho_{\pm})$
under the condition of Eq.~(\ref{eq:condition-WTI})
if $A > 0$ is assumed without loss of generality.
To do so, 
let us examine the two cases of $\gamma M(\rho_{\pm})=iA(\rho_{\pm})$
and $\gamma M(\rho_{\pm})=-iA(\rho_{\pm})$.
In the first case, $\rho_{\pm}$ is obtained as
\begin{align}
     \label{eq:1pm}
  \rho_{1\pm}
  = \frac{\gamma\tilde{m}_{0}
          \pm\sqrt{D}}
         {2\left(\gamma m_{2\parallel}+\frac{A}{2}\right)} ,
\end{align}
while
\begin{align}
     \label{eq:2pm}
  \rho_{2\pm}
  = \frac{\gamma\tilde{m}_{0}
          \pm\sqrt{D}}
         {2\left(\gamma m_{2\parallel}-\frac{A}{2}\right)}
\end{align}
in the second case, where
\begin{align}
   D 
   \equiv (\gamma\tilde{m}_{0})^{2}-4(\gamma m_{2\parallel})^{2}+A^{2} .
\end{align}
We show below that $|\rho_{1\pm}| < 1$ always holds
while $|\rho_{2\pm}| < 1$ never holds.
That is, the appropriate solutions are obtained in the case of
$\gamma M(\rho_{\pm})=iA(\rho_{\pm})$.
We separately consider the cases of $D < 0$ and $D > 0$ below.
Note that $m_{0} < 0$ and $m_{2\parallel} > 0$ are implicitly assumed
in Eq.~(\ref{eq:condition-WTI}).

\subsection{The case of $D < 0$}

In this case, Eqs.~(\ref{eq:1pm}) and (\ref{eq:2pm}) are rewritten as
\begin{align}
  \rho_{1\pm}
  = \frac{\gamma\tilde{m}_{0}
          \pm i \sqrt{-D}}
         {2\left(\gamma m_{2\parallel}+\frac{A}{2}\right)}
  = \rho_{2\pm}^{-1} .
\end{align}
This immediately yields
\begin{align}
  |\rho_{1\pm}|
  = \frac{\left|\gamma m_{2\parallel}-\frac{A}{2}\right|}
         {\left|\gamma m_{2\parallel}+\frac{A}{2}\right|}
  < 1 <
  \frac{\left|\gamma m_{2\parallel}+\frac{A}{2}\right|}
         {\left|\gamma m_{2\parallel}-\frac{A}{2}\right|}
  = |\rho_{2\pm}| .
\end{align}

\subsection{The case of $D > 0$}

In this case, we can show from Eqs.~(\ref{eq:1pm}) and (\ref{eq:2pm})
that $\rho_{1\pm}\rho_{2\mp} = 1$,
and that $|\rho_{2+}| > |\rho_{1+}|$ and $|\rho_{2-}| > |\rho_{1-}|$
since $m_{2\parallel} > 0$ and $A > 0$ are assumed.
Let us separately treat the two cases of $\tilde{m}_{0} > 0$
and $\tilde{m}_{0} < 0$.

If $\tilde{m}_{0} > 0$,
we find that $|\rho_{2+}| > |\rho_{2-}|$ from Eq.~(\ref{eq:2pm}).
The combination of this with $|\rho_{2-}| > |\rho_{1-}|$ yields
\begin{align}
  |\rho_{2+}| > |\rho_{1-}| ,
\end{align}
indicating that $|\rho_{2+}| > 1$ since $\rho_{1-}\rho_{2+} = 1$.
Thus, we see that $|\rho_{2\pm}| < 1$ never holds.
Here, we also find from Eq.~(\ref{eq:1pm}) with $\tilde{m}_{0} > 0$
that $\rho_{1+} > |\rho_{1-}|$.
The above argument indicates that the solution satisfying
the boundary condition can be constructed when
\begin{align}
  1 > \rho_{1+} .
\end{align}
We can show that this always holds under the condition
of Eq.~(\ref{eq:condition-WTI}).

If $\tilde{m}_{0} < 0$,
we find that $|\rho_{2-}| > |\rho_{2+}|$ from Eq.~(\ref{eq:2pm}).
The combination of this with $|\rho_{2+}| > |\rho_{1+}|$ yields
\begin{align}
  |\rho_{2-}| > |\rho_{1+}| ,
\end{align}
indicating that $|\rho_{2-}| > 1$ since $\rho_{1+}\rho_{2-} = 1$.
Thus, we see that $|\rho_{2\pm}| < 1$ never holds.
Here, we also find from Eq.~(\ref{eq:1pm}) with $\tilde{m}_{0} < 0$
that $|\rho_{1-}| > |\rho_{1+}|$.
The above argument indicates that the solution satisfying
the boundary condition can be constructed when
\begin{align}
  1 > |\rho_{1-}| .
\end{align}
We can show that this always holds under the condition
of Eq.~(\ref{eq:condition-WTI}).

\end{document}